# Design and Performance study of Smart Antenna Systems for WIMAX applications

Ayman Abdallah, Chibli Joumaa, Seifedine Kadry

American University oft he Middle East, Egaila, Kuwait
{Ayman.abdallah@aum.edu.kw, Chibli.joumaa@acm.edu.kw,
Seifedine.kadry@aum.edu.kw }

**Abstract.** In this paper we propose an approach that uses homodyne receivers todesign smart antenna systems. The receivers functions are to detect angles of arrivals of seven incoming RF signals using MUSIC or ESPRIT algorithms. The characteristics of each algorithm are critical for the system's precision as well as receivers types. Results are deduced from the simulation of each system, using the Advanced Design System (ADS) and MATLAB. These are compared to results deduced from real systems in the WIMAX (3.5GHz) domains.

**Keywords:** Smart Antennas, DOA, Spatial Smoothing, ESPRIT, MUSIC, Adaptive Antennas.

## 1 Introduction

The problem of localization of sources radiating energy by observing their signal received at spatially separated sensors is of considerable importance, occurring in many fields, including radar, sonar, mobile communications, radio astronomy, and seismology.[1]

Thus a variety of methods, using heterodyne or homodyne techniques, for the Detection Of Arrivals (DOA) estimation are used including spectra estimation, minimum-variance distortion less response estimator, linear prediction, maximum entropy, and maximum likelihood [2]. In addition to previous methods the most famous methods used in DOA are Eigen structure methods, including many versions of MUSIC algorithms [3] [9], minimum norm method [4], ESPRIT method [5] [10], and the weighted subspace fitting method.

The main goal of this paper is to provide a step by step methodology in simulation and in conception of smart antenna systems that use 6-port or 5-port reflectometers [6] [7] [8]. The purpose of the simulation is to compare the quality of the detected angles of both circuits, especially when the number of the incoming RF signals increases till seven. A comparison is also made to a smart antenna system that uses either ESPRIT or MUSIC algorithms. This paper consists of seven main sections.

Section 2 presents a general introduction of smart antennas. Section 3 explains the theory of the MUSIC [9] and ESPRIT [10] algorithms. Whereas section 4 describes the experimental results of the implementation of a smart antenna system using 5-ports reflectometer .Finally the conclusions are summarized in section 5.

## 2  Smart Antennas Systems

Smart antennas are essential parts of the new generation of mobile systems. They can be used to exploit spatial and spectral characteristics of the incoming signals to provide highly accurate location information [1]. Before implementing such systems, a simulation phase must be present in order to optimize its efficiency. A smart antennas system is made of two main parts which are the hardware part and signal processing part. The hardware part is responsible of capturing RF signals and converting them to low frequency (LF) signals. The signal processing part, section three, detects the complex envelopes of the LF signal and provides an estimation of the arrival angles. A physical smart antenna is composed of a set of receiving antenna injects the RF signals into Low Noise Amplifiers (LNA). The issued signals are down-converted to LF signals [6] as shown in Figure 1.

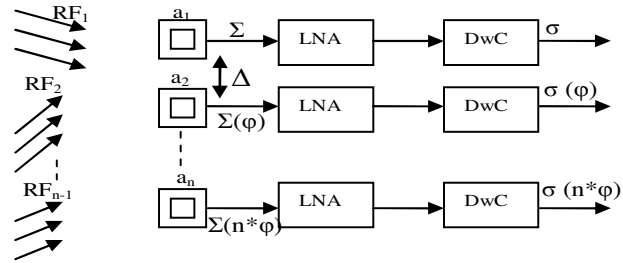

**Fig.1.** Smart antennas hardware system.

Figure1 represents the hardware part of a smart antennas system. "n" coplanar antennas separated by a distance of Δ, transform the incoming electromagnetic signals into electrical ones Σs. Taking a1 as a phase reference, a2 generates an electrical signal from an incoming RFi signal with a phase delay (φ) as shown in the figure below.

The electrical phase delay between two consecutive antennas φio may be computed as $\varphi_i = \dfrac{2\pi \Delta \sin(\theta_i)}{\lambda}$ . $\theta_i$ Is the RF shift angle from the normal of the antennas, λ is the wave length of the central frequency RF.

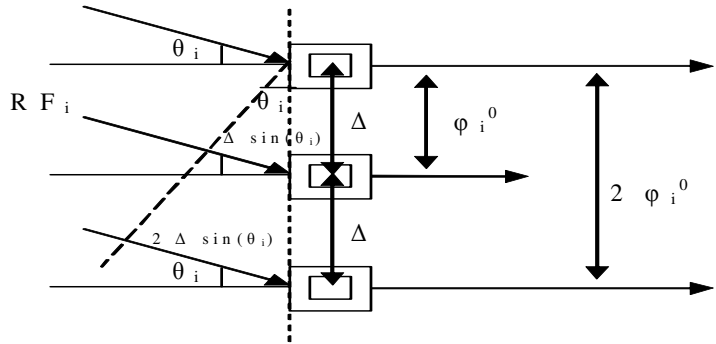

**Fig.2.** Geometry structures in coplanar antennas.

LNAs amplify RF signals and create an additional phase delay. In this paper, we will consider that the software phase delay calculations overcome the later problem by calibrating the down-converters.

Down converters are responsible of transforming RF signals into LF which are easy to manipulate by the software phase. Two major types of down-converter are used in smart antennas, Microwave mixers and Homodyne receivers. In this paper, we will use homodyne receivers and especially six-port reflectometers.

The signal processing phase is dedicated for the Detection Of Arrivals DOA and Signal to Noise Ratios (SNR) determination. The used software programs can be applied in both of real and simulated systems. The smart antenna circuits, including antennas, low noise amplifiers LNA and homodyne receiver junctions are calibrated to provide the required data for the MUSIC and ESPRIT algorithms.

## 3  Smart Antenna in Simulation

In order to stimulate a smart antenna system we utilized the advanced design system (ADS) as a circuit simulation and Matlab tool as signal processor [11]. The characteristics of the smart antenna are derived from the following simulation.

### 3.1  Performance Comparison of MUSIC and ESPRIT

Music & Esprit utilize the $\frac{a_2}{a_1}$ ratio of each reflectometer (5or 6-port) in order to estimate the angle of arrivals. We simulated the smart antenna to detect the following theoretical AOA: 51.10, 19.50, 00, -19.50, -51.10. The result of the simulation is shown in the Table 1.

**Table1.** DOA with MUSIC and ESPRIT

| ESPRIT | | MUSIC | |
|---|---|---|---|
| DOA | S dB | DOA | S dB |
| 51.0 | 53.1 | 51.1 | 21.5 |
| 19.4 | 33.4 | 19.5 | 24.95 |
| -0.0 | 32 | 0 | 25.9 |
| -19.5 | 31.7 | - 19.5 | 23.34 |
| -51.0 | 30.4 | -51.7 | 23.2 |

The above table demonstrates that we can use either MUSIC or ESPRIT algorithms to get a precise result. Note that the simulation was done in a noiseless environment.

### 3.2 Performance Comparison of five and six port reflectometers

In order to compare their performances in DOA, we simulated seven signals to be detected by a smart antenna using six port reflectometers and another one using five ports reflectometers. Both systems perform well as shown in the following figure.

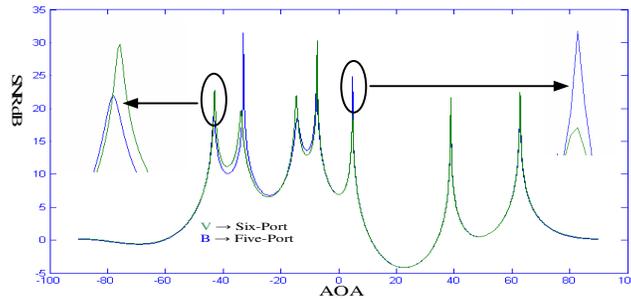

**Fig.7.** DOA of seven RF signals

In the case of instability of the LO power of the five port reflectometer, the results become inaccurate, as shown in Figure 8.

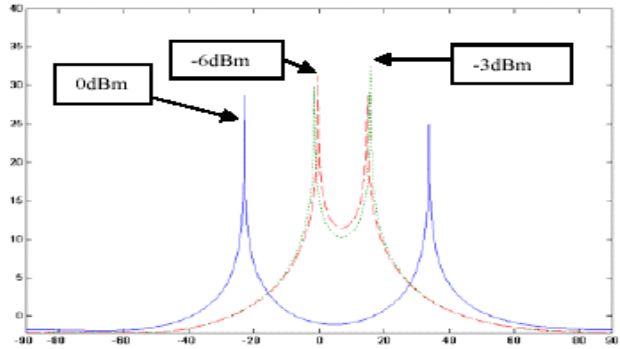

**Fig.8.** DOA of 2 signals (-230 and 340) in instable five ports environments

## 4 Experimental Results

A feasibility study has been made by the conception of a smart antenna demonstrator as shown in figure 9 [12].

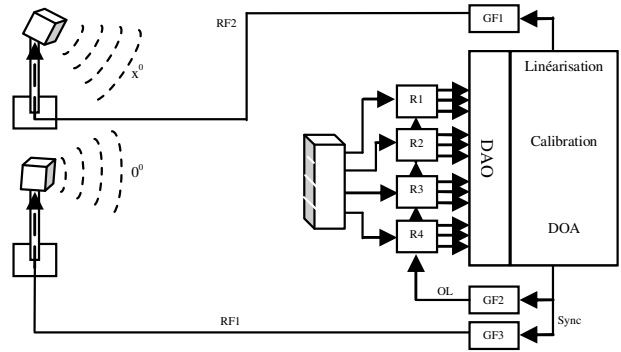

**Fig.9.** Smart antenna system demonstrator.

The smart antenna demonstrator uses 3 RF generators synchronized together representing the LO source and the RF sources conducted to the antenna emitters. The system is calibrated using Neveux's method.

### 4.1 Experimental Comparison of MUSIC and ESPRIT

In order to demonstrate the good performance of the DOA algorithms, MUSIC&ESPRIT, we utilized the complex envelopes provided by the demonstrator.

We generated 2 RF signals of AOA (0° and 34°) uncorrelated in frequency, we deduced the good performance of our algorithm as shown in figure 10.

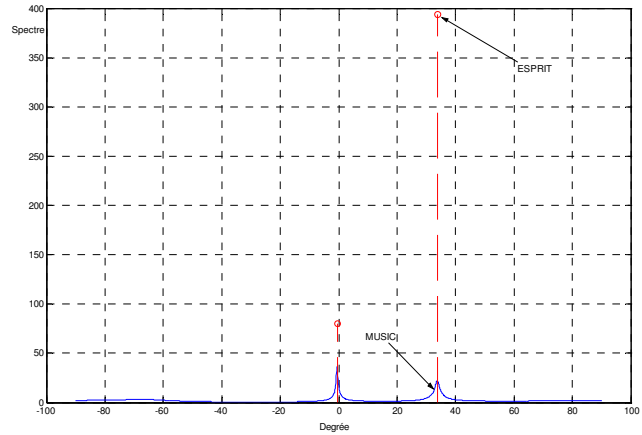

**Fig.10.** DOA of 0° and 34° with MUSIC and ESPRIT

### 4.2 Noisy Environment

The second measurement is made to compare the efficiency of both algorithms in a noisy environment. As shown earlier, both algorithms have a good performance in a noiseless environment. Whereas, in a noisy environment (SNR decreased by 20dB from the initial experiment), we observed a degradation in the performance of the MUSIC algorithm, while the ESPRIT algorithm maintain an acceptable one, as shown in figure 11.

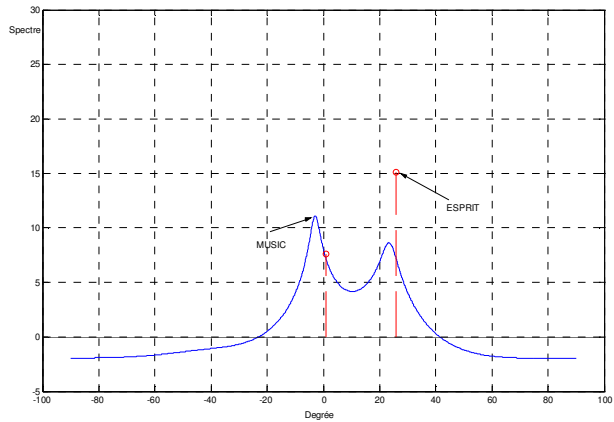

**Fig.11.** Decrease of SNR by 20dB

### 4.3 Resolution

The resolution is calculated by a progressive reduction of the angular difference of the two AOA. In simulation both MUSIC and ESPRIT provide a 10 resolution. The experimental measurements showed that when we had angular separations that are below 50 both algorithm fail, as shown in figure 12.

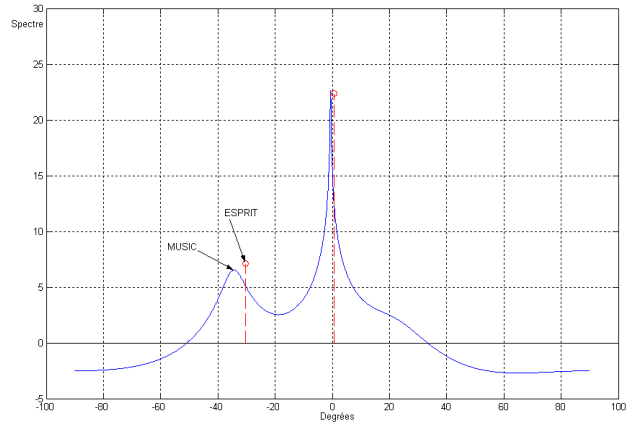

**Fig.12.** DOA of 2 RF separated by less than 50.

### 4.4 Multipath

To overcome the problem of multipath that is present in an environment with high level of signal correlations, we had to pre-treat the signals with a smoothing algorithm. In order to demonstrate the efficiency of this algorithm, we used 2 RF signals with 00 and 270 AOA respectively, having the major power in the 270 signal. The results provided by MUSIC & ESPRIT without preprocessing, demonstrates the failure of the system. The result provided by the system when applying the preprocessor algorithm is shown in the following figure.
interferer RF signal and detected the desired signal.

## 5 Conclusion

This paper gives a clarified idea about simulation and design of smart antenna systems. A performance comparison of two receivers six port and five ports reflectometers was presented. A performance comparison for two DOA algorithms which are ESPRIT and MUSIC was also presented. The homogeneity of the results in simulations and in experiments was demonstrated in terms of accuracy, noise resolution and multipath.

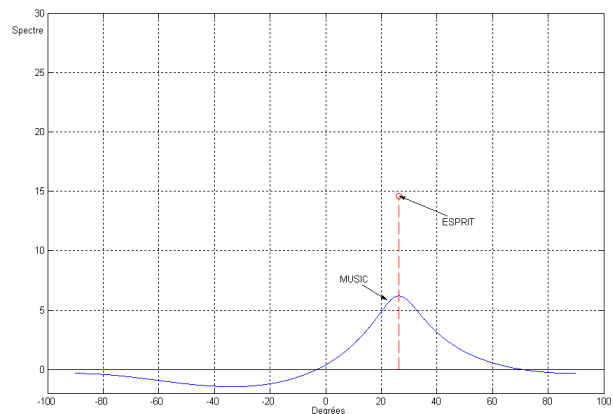

**Fig.13.** Spatial smoothing for multipath rejection.

The above figure shows that the spatial smoothing algorithm has eliminated the

## References

1. Godara, L.C., Application to antenna arrays to mobile communications. Part II: Beamforming and direction of arrival considerations, IEEE Proc., 85, 1195–1247, 1997.
2. Lal Chand Godora, "Smart Antennas", CRC PRESS LLC 2004.
3. R. O. Schmidt, "Multiple emitter location and signal parameter estimation," IEEE Trans. Antenna and Propag., vol. 34,pp. 276-280, 1986.
4. Buckley, K.M. and Xu, X.,L., Spatial spectrum estimation in a location sector, IEEE Trans. Acoust. Speech Signal Process., 38, 1842–1852, 1990.
5. Yuen, N. and Friedlander, B., Asymptotic performance analysis of ESPRIT, higher order ESPRIT, and virtual ESPRIT algorithms, IEEE Trans. Signal Process., 44, 2537–2550, 1996.
6. Amante Garcia, Conception d'un radar d'aide à la conduite automobile utilisant un système discriminateur de fréquence type six-port, Thèse ENST, 2003.
7. Frank Weidmann, Développements pour des applications grands publiques du réflectometre six-portes: algorithme de calibrage robuste, réflectometre à très large bande et réflectometre intégré MMIC. Thesis ENST Paris1997.
8. G.Neveux. demodulateur directe de signaux RF multi-mode et multi-bande utilisant la technique "cinq-port", these ENST Paris 2003.
9. Ildar Urazghildiiev, MUSIC with Outlier Rejection: A New Tool for Improving the Estimation Performance at the Threshold Region of the SNR, European Radar Conference, Amsterdam 2004.
10. Roy, R. and Kailath, T., ESPRIT: estimation of signal parameters via rotational invariance techniques, IEEE Trans. Acoust. Speech Signal Process. 37, 984–995, 1989.
11. Ayman Abdallah, Stability study of smart antenna systems using homodyne receivers for DOA estimation, Wseas 2006.
12. Ayman Abdallah, Contribution à l'étude d'une antenne adaptative et à la conception d'un démonstrateur, thèse XLIM Limoges 2007